\documentclass[journal=nalefd,manuscript=article,layout=twocolumn]{achemso}

\usepackage[version=3]{mhchem}
\usepackage{bm}                      
\usepackage{hyperref}
\usepackage{physics}

\let\oldmaketitle\maketitle
\let\maketitle\relax

\author{Ata Ulhaq}
\email{a.ulhaq@sheffield.ac.uk}
\author{Qingqing Duan}
\affiliation{Department of Physics and Astronomy, University of Sheffield, Sheffield, S3 7RH, United Kingdom.}
\author{Fei Ding}
\author{Eugenio Zallo}
\affiliation{Institute for Integrative Nanoscience, IFW Dresden, Helmholtz str. D-01069, Dresden, Germany.}
\alsoaffiliation{Paul-Drude-Institut f\"{u}r Festk\"{o}rperelektronik, Hausvogteiplatz 5-7, 10117 Berlin, Germany}
\author{Oliver G. Schmidt}
\affiliation{Institute for Integrative Nanoscience, IFW Dresden, Helmholtz str. D-01069, Dresden, Germany.}
\author{Maurice S. Skolnick}
\author{Alexander I. Tartakovskii}
\affiliation{Department of Physics and Astronomy, University of Sheffield, Sheffield, S3 7RH, United Kingdom.}
\author{Evgeny A. Chekhovich}
\email{e.chekhovich@sheffield.ac.uk}
\affiliation{Department of
Physics and Astronomy, University of Sheffield, Sheffield, S3 7RH,
United Kingdom.}

\title[An \textsf{achemso} demo]
  {Electron and nuclear spin properties of the nanohole-filled GaAs/AlGaAs quantum dots}

\begin{document}

\begin{tocentry}

 \includegraphics[bb=0pt 0pt 320pt 130pt, width=9cm]{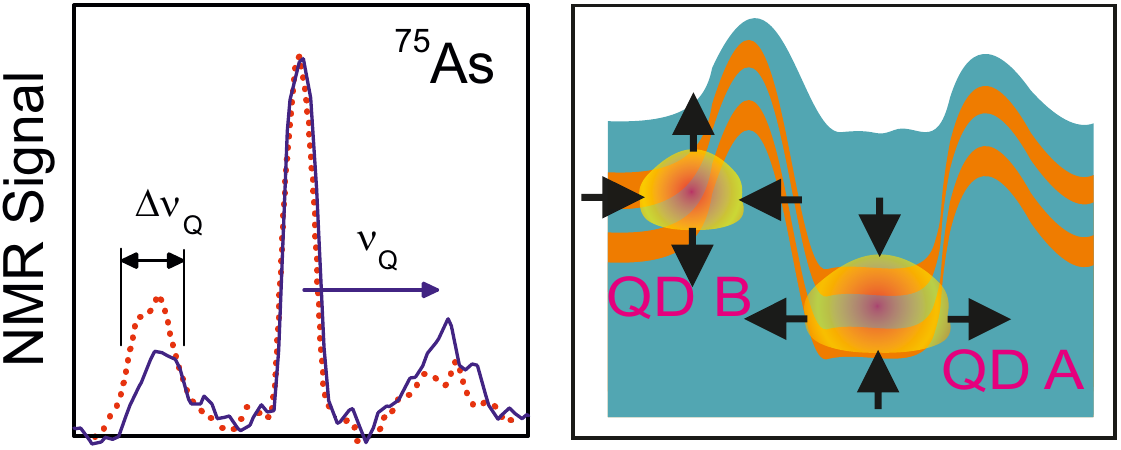}

\end{tocentry}

\twocolumn[
\begin{@twocolumnfalse}
\oldmaketitle
\begin{abstract}
GaAs/AlGaAs quantum dots grown by \textit{in~situ} droplet etching
and nanohole infilling offer a combination of strong charge
confinement, optical efficiency, and spatial symmetry required for
polarization entanglement and spin-photon interface. Here we study
spin properties of such dots. We find nearly vanishing electron
$g$-factor ($g_e<0.05$), providing a route for electrically driven
spin control schemes. Optical manipulation of the nuclear spin
environment is demonstrated with nuclear spin polarization up to
$60\%$ achieved. NMR spectroscopy reveals the structure of two
types of quantum dots and yields the small magnitude of residual
strain $\epsilon_b<0.02\%$ which nevertheless leads to long
nuclear spin lifetimes exceeding 1000~s. The stability of the
nuclear spin environment is advantageous for applications in
quantum information processing.
\end{abstract}
\end{@twocolumnfalse}
]


Central spin in semiconductor quantum dots is a prime candidate
for applications in quantum information technologies
\cite{Kloeffel2013,Gywat2010}. It is relatively isolated from the
solid state effects and at the same time is accessible for
coherent manipulation and can be interfaced optically. The
coherence in this system is mainly limited by the hyperfine
coupling with the nuclear spin bath
\cite{Khaetskii2002,Merkulov2002}. Single spin qubit manipulation
in these structures, therefore, demands an auxiliary control over
nuclear spin environment. Such control can be realized by
maximizing polarization of $10^4-10^5$ nuclei in a single quantum
dot \cite{Coish2004,Bracker2005,Urbaszek2007}, enabling the
formation of well-defined nuclear spin states and in effect
reducing the influence of the nuclear field fluctuations
\cite{Reilly2008,Issler2010}.

Central spin manipulation in semiconductor quantum dot (QD) system
using resonant ultrafast optical pulses
\cite{Press2010,DeGreve2011} has been demonstrated but scalability
in such schemes is challenging. An alternative approach is to
induce controlled spin rotation by manipulating the coupling to
the external magnetic field \cite{Kato2003}. This can be achieved
by electrical modulation of the $g$-factor. However, such scheme
critically depends on the ability to change the sign of $g$, thus
requiring quantum dots with close to zero electron or hole
$g$-factor \cite{Pingenot2008,Pingenot2011}.

Self-assembled InGaAs/GaAs QD has been the primary system of
choice for spin studies over the last two decades, as quantum
confinement in monolayer-fluctuation GaAs/AlGaAs dots is too weak.
Only recently the potential of droplet epitaxial (DE) grown GaAs
QDs has been identified
\cite{Belhadj2008,Abbarchi2010,Sallen2014}. In particular
nanohole-filled droplet epitaxial (NFDE) dots formed by $in~situ$
etching and nanohole infilling \cite{Atkinson2012} provide
confinement and excellent optical efficiency, while on the other
hand exhibiting high symmetry not achievable previously in
self-assembled dots \cite{HuoAPL2013}. Such unique combination of
properties make NFDE dots ideal candidates for polarization
entanglement and spin-photon interface \cite{Gao2015}. This system
has already exhibited an efficient interface between rubidium
atoms and a quantum dot \cite{Akopian2011}. However, the
understanding of the spin properties in such quantum dots is still
lacking.

Here we use optical and nuclear magnetic resonance (NMR)
spectroscopy to study the properties of the single charge spins
and nuclear spin environment in NFDE grown GaAs/AlGaAs QDs.
Magneto-photoluminescence measurements reveal close-to-zero
electron $g$-factor, due to the electron wavefunction overlap with
the AlGaAs barrier. We demonstrate efficient dynamic nuclear
polarization (DNP) as large as $60~\%$. By measuring the
excitation wavelength dependence we identify three mechanisms of
DNP: (i) via optical excitation of the quantum well states, (ii)
via resonant optical excitation of the dot ground or excited
states, and (iii) via resonant excitation of the neighboring dot
made possible by inter-dot charge tunneling. Radio frequency (rf)
excitation is used to measure NMR spectra revealing the presence
of small ($<0.02\%$) residual biaxial strain. Surprisingly, we
observe two sub-ensembles of QDs one with compressive and another
with tensile strain along the growth axis: this allows us to
identify these two types of dots as formed in the nanoholes and at
the rims of the nanoholes respectively. We show that small
residual strain results in very stable nuclear spin bath with
nuclear spin relaxation times $>1000$~s, previously achievable
only in self-assembled dots. The properties of the NFDE quantum
dots revealed in this study make them a favorable system for
electrical spin qubit manipulation with a potential for minimized
decoherence effects from the nuclear spin bath.

Single dot photoluminescence (PL) spectroscopy is performed with a
confocal setup which collects PL at low temperature ($T\approx4.2$
K) from a $\sim1~\mu$m spot. Magnetic fields up to $10$ T along
sample growth axis (Faraday geometry) are employed in this study.
The polarization degree of the nuclear spins is probed by
measuring the hyperfine shifts in the Zeeman splitting of the
quantum dot PL. Nuclear spin polarization and NMR spectroscopy
studies are performed using the methods described in Reference
\cite{ChekhovichNatNano2012}.

\begin{figure}[t]
\centering
\includegraphics[bb=0pt 0pt 450pt 380pt, width=8cm]{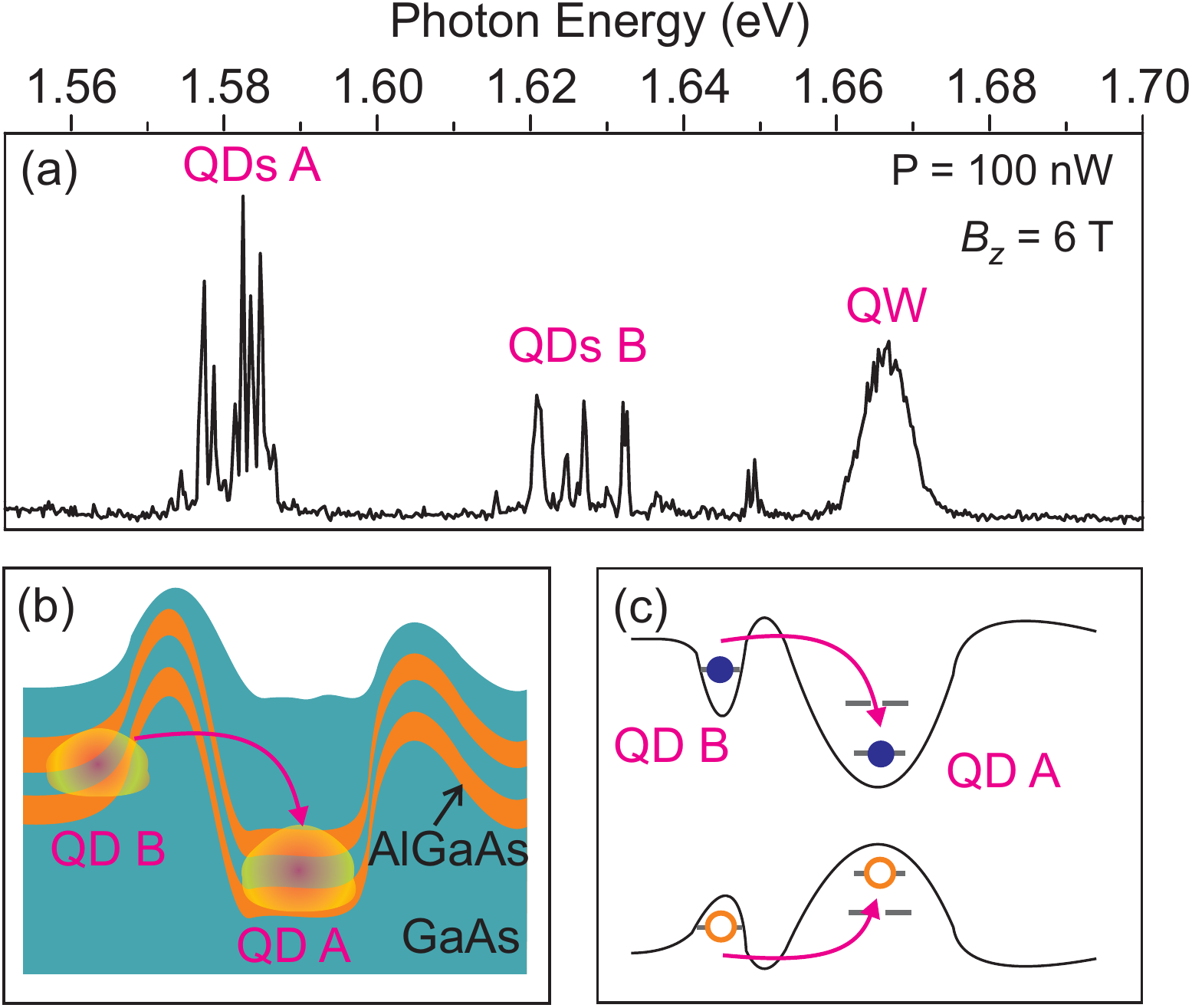}
\caption{(a) Low temperature photoluminescence spectrum showing
emission of two types of quantum dots (A and B) and a quantum well
(QW) measured under non resonant excitation
($E_\textrm{laser}=1.96~$eV) at $B_z=6$~T. (b) Schematic diagram
showing the structure of $in~situ$ nanohole infilled droplet
epitaxial QDs. The deposited GaAs causes formation of a dot inside
the nanohole (QD type A), additional dots (type B) can be formed
at the edge of the nanohole. (c) Schematic bandstructure of a
nanohole infilled dots A and type B. Arrows depict a possible
exciton tunneling from dot B to dot A. } \label{fig1}
\end{figure}


A typical broad PL spectrum of the studied structure under
non-resonant excitation ($E_\textrm{laser}=1.96~$eV) is shown in
Fig. \ref{fig1}(a) at $B_z=~6~$T. Apart from the quantum well (QW)
emission at $E=1.67~$eV, two spectral distributions of QD emission
are observed at $E\approx 1.58~$eV (type A dots) and $E\approx
1.63$eV (type B). The structure of the QD sample used in this
study based on previous AFM measurements \cite{Atkinson2012} is
shown schematically in Fig. \ref{fig1}(b). Inverted pyramid dot
formation is caused by infilling with GaAs of the \textit{in~situ}
etched nanoholes in AlGaAs. These dots are responsible for
emission at $E\approx 1.58~$eV (QDs type A). The topology of the
nanohole allows the formation of smaller dots. As we show based on
NMR spectroscopy measurements such dots are formed at the edges of
the nanohole. Such dots have shallower potential and give rise to
emission at higher energy $E\approx 1.63$eV (QDs type B).

Emission from a QD is a result of recombination of an electron
with spin up $\uparrow$ (or spin down $\downarrow$) and a hole
with spin up $\Uparrow$ (or spin down $\Downarrow$) along $Oz$
axis (parallel to magnetic field $B_z$). An electron-hole pair can
form either a ''bright'' exciton $\ket{\Uparrow\downarrow}$
($\ket{\Downarrow\uparrow}$) with spin projection $+1$ ($-1$) or
optically forbidden ''dark'' exciton \cite{Bayer2002,Poem2010}
$\ket{\Uparrow\uparrow}$ ($\ket{\Downarrow\downarrow}$) with spin
projection $+2$ ($-2$). In QDs with non-ideal symmetry, the
exchange interaction mixes the bright and dark states
\cite{Bayer2002} and hence the dark states gain small oscillator
strength and can be observed in QD PL at low excitation powers.

\begin{figure}[h]
\centering
\includegraphics[bb=71pt 0pt 617pt 570pt, width=8.0cm]{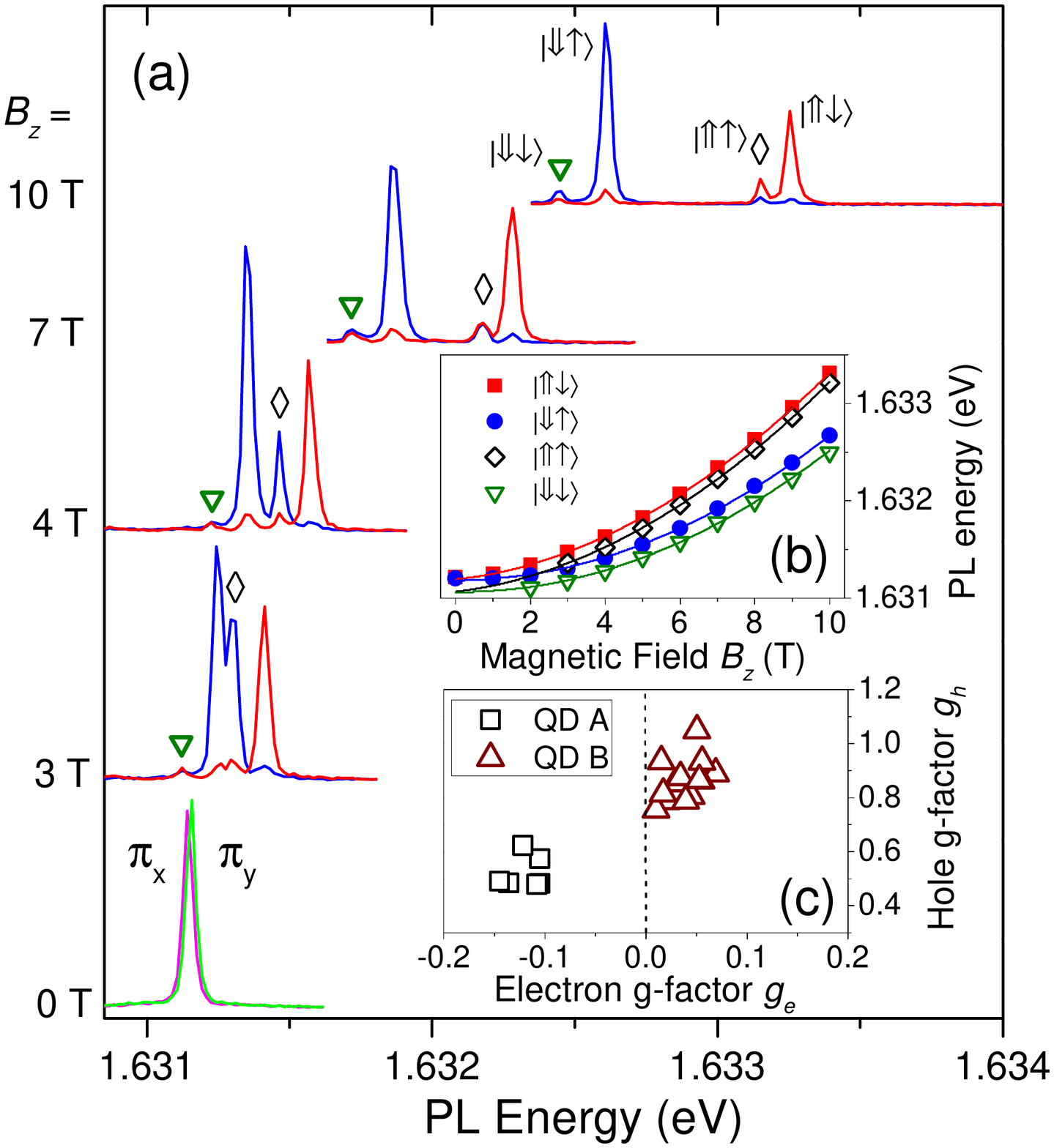}
\caption{(a) Magnetic field $B_z$ dependence of PL emission from
bright and dark excitons in a quantum dot B1 under $\sigma^+$ low
power ($P_{exc}=200~$nW) excitation. Red (blue) lines correspond
to spectra recorded in $\sigma^+$ ($\sigma^-$) polarized
detection, while green and magenta correspond to spectra recorded
in linear polarizations ($\pi_x$, $\pi_y$) at $B_z=0$. The
diamonds ($\diamondsuit$) and triangles ($\bigtriangledown$)
indicate the weak peaks corresponding to $\ket{\Uparrow\uparrow}$
and $\ket{\Downarrow\downarrow}$ dark excitons, respectively. The
two intense peaks correspond to $\ket{\Uparrow\downarrow}$ and
$\ket{\Downarrow\uparrow}$ bright exciton. (b) PL energies of
'dark' (open symbols) and 'bright' (full symbols) exciton peaks
from (a). The solid lines show fit to the data yielding the
electron (hole) g-factor $g_e=0.05$ ($g_h=0.86$) and diamagnetic
shift $\kappa=21.8~\mu\text{eV/T}^2$. (c) Electron ($g_e$) and
hole ($g_h$) $g$-factors measured for several QDs type A (squares)
and QDs B (triangles) from the same sample.} \label{fig2}
\end{figure}

Figure \ref{fig2}(a) presents a series of PL spectra of QD B1
measured at low excitation power and different $B_z$. The emission
of both dark excitons can be observed at finite $B_z$: the
emission lines are marked with $\diamondsuit$ for
$\ket{\Uparrow\uparrow}$ and $\bigtriangledown$ for
$\ket{\Downarrow\downarrow}$ exciton. A finger-print feature of a
''dark'' exciton is its enhanced emission when it anticrosses with
a bright state \cite{Bayer2002}. This is observed in Fig.
\ref{fig2}(a) for the $\ket{\Uparrow\uparrow}$ exciton: at
$B_z=3$~T it has enhanced emission in $\sigma^-$ polarization due
to the mixing with $\ket{\Downarrow\uparrow}$, while at $B_z=10$~T
the enhanced emission in $\sigma^+$ polarization is caused by
mixing with $\ket{\Uparrow\downarrow}$ bright exciton.

The PL peak energies are shown in Fig. \ref{fig2}(b) by the
symbols. The fitting is shown by solid lines. From the fit we find
the electron and hole Land\`{e} $g$-factors along $Oz$ axis $g_e$
and $g_h$ and the diamagnetic shift coefficient $\kappa$. We have
performed magneto-PL measurements for a set of different
individual dots of both type A and B from the same sample. The
extracted electron and hole $g$-factors are plotted in
Fig.~\ref{fig2}(c). Surprisingly, for QDs B, $g_e$ have
close-to-zero values with an average of $g_e\sim+0.05$, an order
of magnitude smaller than for GaAs/AlGaAs QDs formed by natural
fluctuation of the quantum well width
\cite{Puebla2013,GammonPRL2001}. QDs of type A also have small
(and negative g-factors) $g_e\approx -0.1$. The values of
diamagnetic shift $\kappa$ is $16-22~\mu\text{eV}/\text{T}^2$ for
both type A and B dots, which is larger than in natural
GaAs/AlGaAs QDs ($10~\mu\text{eV}/\text{T}^2$, Ref.
\cite{Puebla2013}) and DE-grown GaAs/AlGaAs QDs obtained by
crystallization of Ga droplets ($4-8~\mu\text{eV}/\text{T}^2$,
Ref. \cite{Abbarchi2010}). We attribute large diamagnetic shifts
of the studied NFDE QDs to their larger lateral dimensions
resulting in weaker confinement \cite{Janssens2001}: a typical
nanohole size is $\sim 65~$nm (Ref. \cite{Atkinson2012}) compared
to droplet size $\sim 40~$nm in DE-grown dots \cite{Mano2010}.

Confinement of the charges has also a strong impact on the
Land\`{e} $g$-factors \cite{Snelling1992}. Due to the large
lateral size of the wavefunction in the NFDE dots, the $g$-factor
of the semi-confined electron can be approximated as an average of
the $g$-factors in the dot and in the barrier materials weighted
by the probability of finding the charge in them
\cite{Pryor2006,Hannak1995}. Since electron $g$-factor is negative
in GaAs and positive in AlGaAs \cite{White1972,Kosaka2001} we
ascribe the nearly zero $g_e$ observed in the studied NFDE dots to
significant penetration of the electron wavefunction into the
AlGaAs barrier \cite{Snelling1992,MejiaSalazar2009}. This
conclusion agrees with the observation of smaller $g_e$ in QDs A
[Fig. \ref{fig2}(a)] that have smaller PL energy and hence
stronger exciton confinement.

Based on our observation of very small $g_e\sim0.05$ we expect
that nanohole etching and subsequent infilling process can be
optimized to obtain QDs of either type A or type B with
$g_e\approx0$. The $g_e$ in a QD can be tuned via electric field
\cite{Jovanov2011}. Therefore, adding electrodes to NFDE QD
structures with $g_e\approx0$ would allow coherent rotation with
an access to an arbitrary part of the electron spin Bloch sphere
by switching the value of the electric field
\cite{Kato2003,Bennett2013}. The advantage of this approach is
that a large number of QD spin qubits can be controlled
independently by multiple electrodes on the same semiconductor
chip. This would allow for scalability - the key requirement on
the way for practical implementation of quantum information
processing devices.

\begin{figure*}[t]
\centering
\includegraphics[bb=0pt 130pt 602pt 500pt, width=16cm]{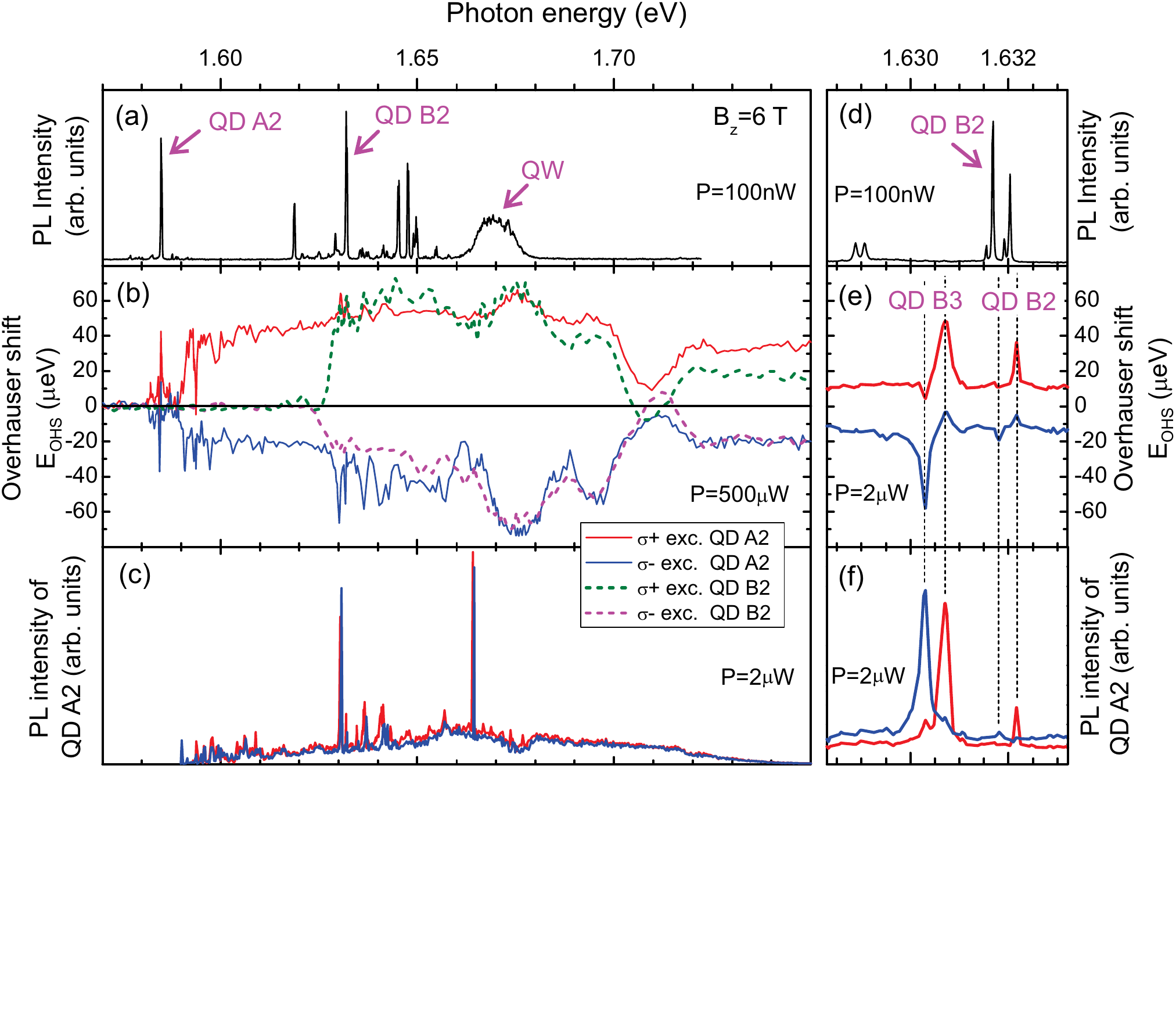}
\caption{Polarization dependent PL, PLE, and DNP spectroscopy on
GaAs QDs at $B_z~=6~$T. (a) PL spectrum showing QW, QD A2 and QD
B2 emission under non resonant excitation
($E_\textrm{laser}=1.96~$eV). (b) Overhauser shift measured on
dots A2 (blue line for $\sigma^-$ and red line for $\sigma^+$
excitation) and B2 (magenta line for $\sigma^-$ and green line for
$\sigma^+$ excitation) as a function of the laser excitation
energy $E_\textrm{laser}$. (c) PL emission magnitude of QD A2 (red
for $\sigma^+$ and blue for $\sigma^-$ excitation) as a function
of $E_\textrm{laser}$. (d) High resolution PL spectrum of QD B2.
(e) Low power DNP measured on QD A2 as $E_\textrm{laser}$ is
scanned close to QD B2 resonance. (f) Zoomed-in view of (c).}
\label{fig3}
\end{figure*}

Since in III-V semiconductors the electron spin is coupled to the
nuclear spin environment via hyperfine interaction, it is
important to understand the properties of the nuclear spin bath
and establish the techniques for its manipulation. To monitor the
polarization of the QD nuclei we measure the splitting $\Delta
E_{\ket{\Uparrow\downarrow},\ket{\Downarrow\uparrow}}$ between the
Zeeman components ($\ket{\Uparrow\downarrow}$ and
$\ket{\Downarrow\uparrow}$) of the bright exciton. The Overhauser
shift $E_{\textrm{OHS}}$ is the change in $\Delta
E_{\ket{\Uparrow\downarrow},\ket{\Downarrow\uparrow}}$ and
characterizes the degree of nuclear spin polarization.

We start by investigating the dynamic nuclear polarization (DNP)
under optical pumping, in particular the dependence on the energy
of the laser excitation $E_\textrm{laser}$. Fig. \ref{fig3}(a)
shows PL spectrum of QD A2 used in these experiments; the emission
from the type B dots is also observed. Photoluminescence
excitation (PLE) spectrum of QD A2 under excitations with both
circular polarizations at low optical power of 2~$\mu$W is
presented in Fig. \ref{fig3}(c). The PLE data reveals sharp peaks
for $E_\textrm{laser}$ up to $\sim1.61$~eV, $\sim25$~meV above QD
A2 ground state energy $\sim1.585$~eV -- these can be ascribed to
the excited states of the QD A2. Above
$E_\textrm{laser}\sim1.61$~eV the PLE trace has a broad
background. We attribute this to the large lateral size of the
type A quantum dot resulting in high spectral density of the
excited states merging into a continuum. However, in addition to
this broad background there is a set of sharp PLE peaks observed
above $E_\textrm{laser}=1.63~$eV. These have energies close to the
energies of the PL peaks of the type B quantum dots, suggesting
that there is an efficient mechanism for injecting electron-hole
pairs into A dots via the excitation of the B dots as shown
schematically in Fig.~1(c).

In addition to PLE spectroscopy we have measured the nuclear
polarization $E_{\textrm{OHS}}$ as a function of
$E_\textrm{laser}$. The red (blue) line in Fig. \ref{fig3}(b)
shows positive (negative) $E_{\textrm{OHS}}$ induced in QD A2
under $\sigma^+$ ($\sigma^-$) polarized high power
($P=500$~$\mu$W) optical excitation. (The measurement was
performed with pump-probe techniques\cite{ChekhovichNatNano2012}).
The results allow to identify at least three mechanisms of DNP:
(i) It can be seen that the highest efficiency DNP with
$|E_{\textrm{OHS}}|\geq$70~$\mu$eV is achieved for
$E_\textrm{laser}\sim1.675$~eV corresponding to the QW states -
this is similar to DNP via QW states in fluctuation GaAs quantum
dots \cite{Bracker2005,Nikolaenko2009} and DNP via the wetting
layer states in self-assembled dots
\cite{Urbaszek2007,Puebla2013}. (ii) A series of sharp peaks
between $E_\textrm{laser}=1.585 - 1.60$~eV is observed correlated
with the PLE peaks in Fig. \ref{fig3}(c). These correspond to DNP
via resonant optical excitation either of the QD A2 ground state
or excited states (e.g. $p$-shell). Such mechanism is also well
known from the studies on self-assembled quantum dots
\cite{Lai2006,Latta2009,Kloeffel2011,Hoegele2012}. Similar to the
case of PLE, the non-zero background $E_{\textrm{OHS}}$ at all
$E_\textrm{laser}>1.60$~eV is ascribed to nearly continuum
spectrum of the excited states of the NFDE QDs with large lateral
dimensions. (iii) Finally, a set of sharp peaks is observed in
Fig. \ref{fig3}(b) at $E_\textrm{laser}=1.63 - 1.66$~eV. These
peaks are strongly correlated to both PLE and PL peaks of the type
B QDs, suggesting that DNP in one dot (of type A) can be produced
by optical excitation of another dot (of type B). Such mechanism
has not been reported previously and is unique to the NFDE QDs.

In order to understand the mechanism of the inter-dot DNP we
perform high-resolution spectroscopy as shown in Figs.
\ref{fig3}(d-f) where we focus on the range of energies around QD
B2 ground excitonic state. Vertical dashed lines show that with
high accuracy there is a direct correspondence between the peaks
in PLE (f) and DNP (e) spectra, confirming that the DNP in QD A2
is a result of the resonant optical electron-hole injection into
the dot. One doublet of the circularly polarized PLE and DNP peaks
at $1.632~$eV can be attributed to the Zeeman doublet of the QD B2
observed in PL (d). This allows to explain the mechanism of the
inter-dot DNP: under resonant optical excitation an exciton is
generated in QD B2. With a finite probability this exciton can
tunnel into QD A2, where it can exchange electron spin with a
nucleus [resulting in a DNP peak in Fig. \ref{fig3}(e)] and then
recombine [resulting in a PLE peak in Fig. \ref{fig3}(f)]. We note
that the PL peaks of QD B2 in Fig. \ref{fig3}(d) are red-shifted
by $150~\mu$eV from the corresponding PLE and DNP peaks in Figs.
\ref{fig3}(e,f). We ascribe this to Pauli blockade
\cite{Kessler2012}: the PLE absorption in QD B2 is observed only
when QD A2 is empty, by contrast the PL from QD B2 can only be
observed when QD A2 is occupied with an exciton, preventing
further exciton tunneling from QD B2 as well as shifting the
ground state energy of QD B2.

On the other hand, the much stronger PLE and DNP doublets at
$1.630$~eV seemingly have no PL lines from type B QD related to
them. This however, can be understood if we assume that such QD
(that we denote as B3) exists but has much larger tunneling rate
compared to QD B2. Thus the excitons from QD B3 tunnel into QD A2
before they can recombine, as a result the PL from QD B3 is
suppressed while PL and DNP in QD A2 are enhanced. We have also
measured the DNP in QD B2 as a function of $E_\textrm{laser}$ as
shown in Fig.~2(b) by the dashed lines. Importantly there is no
DNP in QD B2 when exciting QD A2: we thus conclude that the
nuclear spin diffusion between the dots is negligible and the
inter-dot DNP in QD A2 under resonant optical excitation of QD B2
is indeed due to the tunneling of the excitons.

Upon examining several QDs from the same sample we found that the
results presented in Fig.~\ref{fig3} are well reproduced in other
dots. The DNP in dots type A induced via it's resonant pumping is
found to be as large as $|E_{\textrm{OHS}}|\approx50~\mu$eV. The
DNP induced via optical pumping into the QW or via tunneling from
type B dots is noticeably larger
$|E_{\textrm{OHS}}|\approx85~\mu$eV, corresponding to polarization
degrees of $\sim60\%$. However, the DNP via inter-dot tunneling
may have an advantage since it allows for a selective control of
the nuclear polarization in individual dots, while non-resonant
excitation of the QW polarizes nuclei in all dots within the laser
spot\cite{Urbaszek2013}.

\begin{figure}[t]
\centering
\includegraphics[bb=0pt 0pt 440pt 360pt, width=8.0cm]{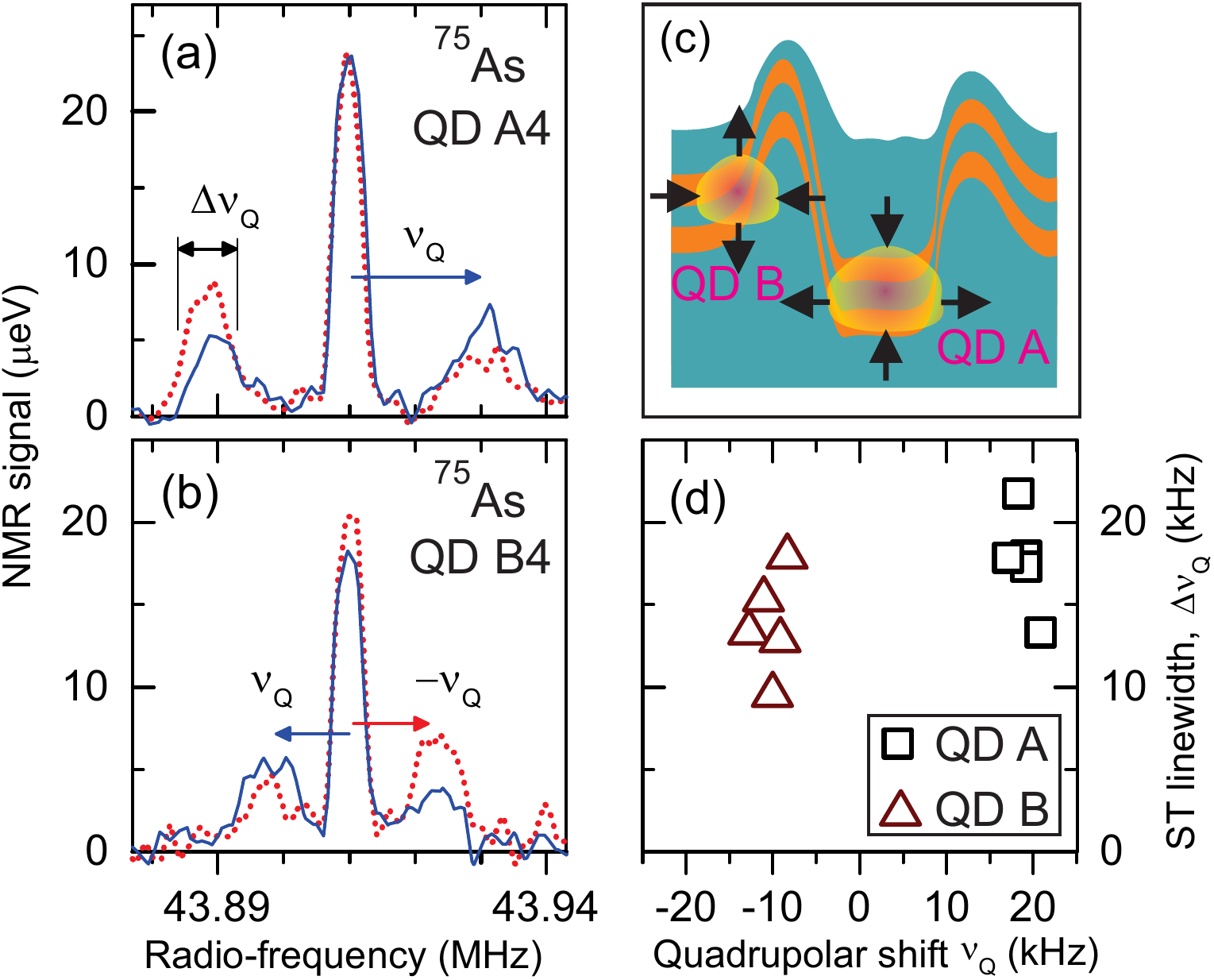}
\caption{Nuclear magnetic resonance spectrum of $^{75}$As nuclei
measured on QD A4 (a) and QD B4 (b) under $\sigma^+$ (red lines)
and $\sigma^-$ (blue lines) optical nuclear spin pumping. The
satellite transitions (STs) are separted from the central
transition (CT) by the strain induced quadrupolar shift $\nu_Q$.
The central transition is resolution limited while the ST width is
$\Delta\nu_Q$. (c) Schematic showing the strain profile in dots A
and B with black arrows indicating the strain directions for both
dots as deduced from the NMR spectra in (a) and (b). (d) Mean
quadrupolar shifts $\nu_Q$ and ST broadening $\Delta\nu_Q$
measured for several QDs type A (squares) and B (triangles).}
\label{fig4}
\end{figure}

The ability to induce large DNP allows us to perform nuclear
magnetic resonance (NMR) spectroscopy in order to investigate the
QD structural properties \cite{ChekhovichNatNano2012}. Figs.
\ref{fig4}(a) and (b) show NMR spectra of the $^{75}$As spin
$I=3/2$ isotope for QDs A4 and B4, respectively. The spectra
contain a narrow (resolution limited) central peak corresponding
to the nuclear spin $-1/2 \leftrightarrow +1/2$ central transition
(CT). Two satellite transitions (STs) $\pm 3/2 \leftrightarrow \pm
1/2$ are shifted by frequency $\mp\nu_Q$ from the CT. The non-zero
$\nu_Q$ reveals the presence of a biaxial elastic strain, even
though the GaAs/AlGaAs structures are expected to be nearly
lattice-matched \cite{Belhadj2008}.

In order to quantify the strain in QDs we first note an asymmetry
observed in the NMR spectra of Figs. \ref{fig4}(a,b): under the
$\sigma^+$ optical pumping the low-frequency (high-frequency) ST
of QD A4 (B4) has increased amplitude. As $\sigma^+$ light
enhances the NMR signal of the $-3/2 \leftrightarrow -1/2$ ST
\cite{ChekhovichNatNano2012}, we conclude that QDs A4 and B4 have
opposite signs of the quadrupolar shifts: $\nu_Q>0$ for QD A4 and
$\nu_Q<0$ for QD B4. NMR measurements on several individual dots
shown in Fig. \ref{fig4}(d) reveal systematic positive values
$\nu_Q\approx+20$~kHz for dots type A (squares) and negative
values $\nu_Q\approx-10$~kHz for dots type B (triangles). The ST
half-widths $\Delta \nu_Q$, reflecting the inhomogeneous
distribution of $\nu_Q$ within the dot are found to vary in the
range $\Delta \nu_Q\sim10-20$~kHz.

Pure hydrostatic strain does not cause nuclear quadrupolar shifts.
But under a uniaxial strain of magnitude $\epsilon_b$ and with
major axis parallel to magnetic field (along $Oz$), the
quadrupolar shift reads as:
\begin{equation}
\nu_Q = \frac{3eQS_{11}\epsilon_b}{2hI(2I-1)},\label{eq:nuQ}
\end{equation}
where $Q$ is the nuclear quadrupolar moment ($\approx
0.31\times10^{-28}~$m$^2$ for $^{75}$As),
$\left|S_{11}\right|\approx 3.9\times 10^{22}$~V/m$^2$ is the
gradient elastic tensor for $^{75}$As in bulk GaAs (the sign of
$S_{11}$ is undefined) \cite{Sundfors1974}, $e$ is electron
charge, and $h$ is the Planck's constant. Thus the average NMR
frequency shift $\nu_Q$ provides a direct measure of the average
strain, while ST linewidth $\Delta \nu_Q$ gives a measure of
strain distribution within the quantum dot.

In disk-shaped (large lateral and small vertical dimensions)
self-assembled InGaAs/GaAs quantum dots the biaxial strain is
positive (tensile along $Oz$ axis) \cite{Pearson2000}. In such
dots negative $\nu_Q$ was found for $^{75}$As nuclei, implying
$S_{11}<0$. For GaAs/AlGaAs NFDE QDs type A we find positive
$\nu_Q$, hence the strain derived from Eq. 1 is negative
(compressive along $Oz$) $\epsilon_b = -0.014 \%$. This, however,
is expected for disk-shaped dots since GaAs lattice constant is
smaller than that of AlGaAs (as opposed to InGaAs/GaAs pair). By
contrast, for QDs type B we find anomalous positive (tensile along
$Oz$) $\epsilon_b = 0.007 \%$. Most importantly, we always observe
either distinctly positive (QDs type B) or distinctly negative
(QDs type A) $\epsilon_b$. This allows us to conclude that there
is no significant overlap between the excitonic wavefunctions in
dots type A and B, as such overlap would have resulted in gradual
transition between tensile and compressive strains leading to
large inhomogeneous broadening of NMR spectra (as observed in
self-assembled dots \cite{ChekhovichNatNano2012}).

The NMR data can be explained consistently if we assume the
structure of the dots is as it is shown in Fig. 1(b) and in Fig.
4(c): while QDs type A are formed by in-filling of the nanohole,
the ''mounds'' formed at the rim of the nanohole (and previously
observed in AFM \cite{Atkinson2012}) create additional confinement
potential resulting in formation of QDs type B. The tensile strain
in QDs type B can be explained by the ''sloped'' AlGaAs barriers
resulting in compressive in-plane strain of GaAs as shown with
arrows in Fig. 4(c). By contrast the topology of QDs type A in the
nanohole is closer to that of a quantum well, so that both AlGaAs
barriers act to stretch the GaAs layer in the horizontal plane
resulting in compressive strain along $Oz$ ($\epsilon_b<0$).
However, estimating the strain in an ideal
GaAs/Al$_{0.4}$Ga$_{0.6}$As quantum well \cite{Grundmann1995} at
low temperature gives much lower $\epsilon_b=-0.1\%$. This
suggests that in QDs A and B there is a significant degree of
in-plane compression resulting most likely from the concave shapes
of the dots.

Despite it's small magnitude the residual strain has a major
impact on the nuclear spin system. From the measurements of the
nuclear spin depolarization in the dark, we observe decay times of
up to 1000~s in both QDs type A and B. This is significantly
longer than in natural fluctuation GaAs/AlGaAs dots where decay
times of $\sim40$~s were found \cite{Nikolaenko2009}, and is
comparable to the decay times in self-assembled InP \cite{InPDyn}
and InGaAs \cite{Maletinsky2009} QDs. The enhanced stability of
the nuclear spin polarization in the NFDE dots can be understood
from the NMR spectra in Figs. \ref{fig4}(a,b): unlike in
fluctuation dots \cite{GammonScience,MakhoninPRB}, the ST and CT
transitions are well resolved, so that the spin exchange between
the adjacent nuclei is significantly inhibited. As a result
nuclear spin diffusion out of the dot is suppressed providing
excellent stability of the nuclear spin magnetization, crucial for
achieving long electron and hole spin coherence but found
previously only in highly strained self-assembled dots.

In conclusion, we have explored $in~situ$ nanohole infilled
droplet epitaxial quantum dot system with respect to electron and
hole spin properties and nuclear spin environment. Investigations
into the Land\`{e} $\textit{g}$-factors have demonstrated a
quantum dot system with electron g-factor $g_{e}$ approaching
zero, making it an exciting platform for fully-scalable electric
spin control schemes based on $\textit{g}$-factor manipulation.
Nuclear spin bath can be manipulated optically with polarization
degrees as large as $60~\%$ achieved reliably. Structural analysis
using NMR spectroscopy reveals small residual strain that switches
from tensile to compressive depending on the type of the dot and
leads to very long nuclear spin lifetimes providing a stable spin
bath environment for the electron or hole spin.

\begin{acknowledgement}
The authors are grateful to Armando Rastelli and Yongheng Huo
(Linz) for the fruitful discussions. This work has been supported
by the EPSRC Programme Grant EP/J007544/1 and the Royal Society.
E.A.C. was supported by a University of Sheffield
Vice-Chancellor's Fellowship.
\end{acknowledgement}




\begin{mcitethebibliography}{51}
\providecommand*\natexlab[1]{#1}
\providecommand*\mciteSetBstSublistMode[1]{}
\providecommand*\mciteSetBstMaxWidthForm[2]{}
\providecommand*\mciteBstWouldAddEndPuncttrue
  {\def\EndOfBibitem{\unskip.}}
\providecommand*\mciteBstWouldAddEndPunctfalse
  {\let\EndOfBibitem\relax}
\providecommand*\mciteSetBstMidEndSepPunct[3]{}
\providecommand*\mciteSetBstSublistLabelBeginEnd[3]{}
\providecommand*\EndOfBibitem{} \mciteSetBstSublistMode{f}
\mciteSetBstMaxWidthForm{subitem}{(\alph{mcitesubitemcount})}
\mciteSetBstSublistLabelBeginEnd
  {\mcitemaxwidthsubitemform\space}
  {\relax}
  {\relax}

\bibitem[Kloeffel and Loss(2013)Kloeffel, and Loss]{Kloeffel2013}
Kloeffel,~C.; Loss,~D. \emph{Annual Review of Condensed Matter
Physics}
  \textbf{2013}, \emph{4}, 51--81\relax
\mciteBstWouldAddEndPuncttrue
\mciteSetBstMidEndSepPunct{\mcitedefaultmidpunct}
{\mcitedefaultendpunct}{\mcitedefaultseppunct}\relax \EndOfBibitem
\bibitem[Gywat \latin{et~al.}(2010)Gywat, Krenner, and Berezovsky]{Gywat2010}
Gywat,~O.; Krenner,~H.~J.; Berezovsky,~J. \emph{Spins in Optically
Active
  Quantum Dots}, 1st ed.; Wiley-VCH, 2010\relax
\mciteBstWouldAddEndPuncttrue
\mciteSetBstMidEndSepPunct{\mcitedefaultmidpunct}
{\mcitedefaultendpunct}{\mcitedefaultseppunct}\relax \EndOfBibitem
\bibitem[Khaetskii \latin{et~al.}(2002)Khaetskii, Loss, and
  Glazman]{Khaetskii2002}
Khaetskii,~A.~V.; Loss,~D.; Glazman,~L. \emph{Phys. Rev. Lett.}
\textbf{2002},
  \emph{88}, 186802\relax
\mciteBstWouldAddEndPuncttrue
\mciteSetBstMidEndSepPunct{\mcitedefaultmidpunct}
{\mcitedefaultendpunct}{\mcitedefaultseppunct}\relax \EndOfBibitem
\bibitem[Merkulov \latin{et~al.}(2002)Merkulov, Efros, and Rosen]{Merkulov2002}
Merkulov,~I.~A.; Efros,~A.~L.; Rosen,~M. \emph{Phys. Rev. B}
\textbf{2002},
  \emph{65}, 205309\relax
\mciteBstWouldAddEndPuncttrue
\mciteSetBstMidEndSepPunct{\mcitedefaultmidpunct}
{\mcitedefaultendpunct}{\mcitedefaultseppunct}\relax \EndOfBibitem
\bibitem[Coish and Loss(2004)Coish, and Loss]{Coish2004}
Coish,~W.~A.; Loss,~D. \emph{Phys. Rev. B} \textbf{2004},
\emph{70},
  195340\relax
\mciteBstWouldAddEndPuncttrue
\mciteSetBstMidEndSepPunct{\mcitedefaultmidpunct}
{\mcitedefaultendpunct}{\mcitedefaultseppunct}\relax \EndOfBibitem
\bibitem[Bracker \latin{et~al.}(2005)Bracker, Stinaff, Gammon, Ware, Tischler,
  Shabaev, Efros, Park, Gershoni, Korenev, and Merkulov]{Bracker2005}
Bracker,~A.~S.; Stinaff,~E.~A.; Gammon,~D.; Ware,~M.~E.;
Tischler,~J.~G.;
  Shabaev,~A.; Efros,~A.~L.; Park,~D.; Gershoni,~D.; Korenev,~V.~L.;
  Merkulov,~I.~A. \emph{Phys. Rev. Lett.} \textbf{2005}, \emph{94},
  047402\relax
\mciteBstWouldAddEndPuncttrue
\mciteSetBstMidEndSepPunct{\mcitedefaultmidpunct}
{\mcitedefaultendpunct}{\mcitedefaultseppunct}\relax \EndOfBibitem
\bibitem[Urbaszek \latin{et~al.}(2007)Urbaszek, Braun, Amand, Krebs, Belhadj,
  Lema\'itre, Voisin, and Marie]{Urbaszek2007}
Urbaszek,~B.; Braun,~P.-F.; Amand,~T.; Krebs,~O.; Belhadj,~T.;
Lema\'itre,~A.;
  Voisin,~P.; Marie,~X. \emph{Phys. Rev. B} \textbf{2007}, \emph{76},
  201301\relax
\mciteBstWouldAddEndPuncttrue
\mciteSetBstMidEndSepPunct{\mcitedefaultmidpunct}
{\mcitedefaultendpunct}{\mcitedefaultseppunct}\relax \EndOfBibitem
\bibitem[Reilly \latin{et~al.}(2008)Reilly, Taylor, Petta, Marcus, Hanson, and
  Gossard]{Reilly2008}
Reilly,~D.~J.; Taylor,~J.~M.; Petta,~J.~R.; Marcus,~C.~M.;
Hanson,~M.~P.;
  Gossard,~A.~C. \emph{Science} \textbf{2008}, \emph{321}, 817--821\relax
\mciteBstWouldAddEndPuncttrue
\mciteSetBstMidEndSepPunct{\mcitedefaultmidpunct}
{\mcitedefaultendpunct}{\mcitedefaultseppunct}\relax \EndOfBibitem
\bibitem[Issler \latin{et~al.}(2010)Issler, Kessler, Giedke, Yelin, Cirac,
  Lukin, and Imamoglu]{Issler2010}
Issler,~M.; Kessler,~E.~M.; Giedke,~G.; Yelin,~S.; Cirac,~I.;
Lukin,~M.~D.;
  Imamoglu,~A. \emph{Phys. Rev. Lett.} \textbf{2010}, \emph{105}, 267202\relax
\mciteBstWouldAddEndPuncttrue
\mciteSetBstMidEndSepPunct{\mcitedefaultmidpunct}
{\mcitedefaultendpunct}{\mcitedefaultseppunct}\relax \EndOfBibitem
\bibitem[Press \latin{et~al.}(2010)Press, De~Greve, McMahon, Ladd, Friess,
  Schneider, Kamp, H\"{o}fling, Forchel, and Yamamoto]{Press2010}
Press,~D.; De~Greve,~K.; McMahon,~P.~L.; Ladd,~T.~D.; Friess,~B.;
  Schneider,~C.; Kamp,~M.; H\"{o}fling,~S.; Forchel,~A.; Yamamoto,~Y. \emph{Nat
  Photon} \textbf{2010}, \emph{4}, 367--370\relax
\mciteBstWouldAddEndPuncttrue
\mciteSetBstMidEndSepPunct{\mcitedefaultmidpunct}
{\mcitedefaultendpunct}{\mcitedefaultseppunct}\relax \EndOfBibitem
\bibitem[De~Greve \latin{et~al.}(2011)De~Greve, McMahon, Press, Ladd, Bisping,
  Schneider, Kamp, Worschech, H\"{o}fling, Forchel, and Yamamoto]{DeGreve2011}
De~Greve,~K.; McMahon,~P.~L.; Press,~D.; Ladd,~T.~D.; Bisping,~D.;
  Schneider,~C.; Kamp,~M.; Worschech,~L.; H\"{o}fling,~S.; Forchel,~A.;
  Yamamoto,~Y. \emph{Nat Phys} \textbf{2011}, \emph{7}, 872 -- 878\relax
\mciteBstWouldAddEndPuncttrue
\mciteSetBstMidEndSepPunct{\mcitedefaultmidpunct}
{\mcitedefaultendpunct}{\mcitedefaultseppunct}\relax \EndOfBibitem
\bibitem[Kato \latin{et~al.}(2003)Kato, Myers, Driscoll, Gossard, Levy, and
  Awschalom]{Kato2003}
Kato,~Y.; Myers,~R.~C.; Driscoll,~D.~C.; Gossard,~A.~C.; Levy,~J.;
  Awschalom,~D.~D. \emph{Science} \textbf{2003}, \emph{299}, 1201--1204\relax
\mciteBstWouldAddEndPuncttrue
\mciteSetBstMidEndSepPunct{\mcitedefaultmidpunct}
{\mcitedefaultendpunct}{\mcitedefaultseppunct}\relax \EndOfBibitem
\bibitem[Pingenot \latin{et~al.}(2008)Pingenot, Pryor, and
  Flatt\'{e}]{Pingenot2008}
Pingenot,~J.; Pryor,~C.~E.; Flatt\'{e},~M.~E. \emph{Applied
Physics Letters}
  \textbf{2008}, \emph{92}, 222502\relax
\mciteBstWouldAddEndPuncttrue
\mciteSetBstMidEndSepPunct{\mcitedefaultmidpunct}
{\mcitedefaultendpunct}{\mcitedefaultseppunct}\relax \EndOfBibitem
\bibitem[Pingenot \latin{et~al.}(2011)Pingenot, Pryor, and
  Flatt\'e]{Pingenot2011}
Pingenot,~J.; Pryor,~C.~E.; Flatt\'e,~M.~E. \emph{Phys. Rev. B}
\textbf{2011},
  \emph{84}, 195403\relax
\mciteBstWouldAddEndPuncttrue
\mciteSetBstMidEndSepPunct{\mcitedefaultmidpunct}
{\mcitedefaultendpunct}{\mcitedefaultseppunct}\relax \EndOfBibitem
\bibitem[Belhadj \latin{et~al.}(2008)Belhadj, Kuroda, Simon, Amand, Mano,
  Sakoda, Koguchi, Marie, and Urbaszek]{Belhadj2008}
Belhadj,~T.; Kuroda,~T.; Simon,~C.-M.; Amand,~T.; Mano,~T.;
Sakoda,~K.;
  Koguchi,~N.; Marie,~X.; Urbaszek,~B. \emph{Phys. Rev. B} \textbf{2008},
  \emph{78}, 205325\relax
\mciteBstWouldAddEndPuncttrue
\mciteSetBstMidEndSepPunct{\mcitedefaultmidpunct}
{\mcitedefaultendpunct}{\mcitedefaultseppunct}\relax \EndOfBibitem
\bibitem[Abbarchi \latin{et~al.}(2010)Abbarchi, Kuroda, Mano, Sakoda, and
  Gurioli]{Abbarchi2010}
Abbarchi,~M.; Kuroda,~T.; Mano,~T.; Sakoda,~K.; Gurioli,~M.
\emph{Phys. Rev. B}
  \textbf{2010}, \emph{81}, 035334\relax
\mciteBstWouldAddEndPuncttrue
\mciteSetBstMidEndSepPunct{\mcitedefaultmidpunct}
{\mcitedefaultendpunct}{\mcitedefaultseppunct}\relax \EndOfBibitem
\bibitem[Sallen \latin{et~al.}(2014)Sallen, Kunz, Amand, Bouet, Kuroda, Mano,
  Paget, Krebs, Marie, Sakoda, and Urbaszek]{Sallen2014}
Sallen,~G.; Kunz,~S.; Amand,~T.; Bouet,~L.; Kuroda,~T.; Mano,~T.;
Paget,~D.;
  Krebs,~O.; Marie,~X.; Sakoda,~K.; Urbaszek,~B. \emph{Nat Commun}
  \textbf{2014}, \emph{5}, 3268\relax
\mciteBstWouldAddEndPuncttrue
\mciteSetBstMidEndSepPunct{\mcitedefaultmidpunct}
{\mcitedefaultendpunct}{\mcitedefaultseppunct}\relax \EndOfBibitem
\bibitem[Atkinson \latin{et~al.}(2012)Atkinson, Zallo, and
  Schmidt]{Atkinson2012}
Atkinson,~P.; Zallo,~E.; Schmidt,~O.~G. \emph{Journal of Applied
Physics}
  \textbf{2012}, \emph{112}, 054303\relax
\mciteBstWouldAddEndPuncttrue
\mciteSetBstMidEndSepPunct{\mcitedefaultmidpunct}
{\mcitedefaultendpunct}{\mcitedefaultseppunct}\relax \EndOfBibitem
\bibitem[Huo \latin{et~al.}(2013)Huo, Rastelli, and Schmidt]{HuoAPL2013}
Huo,~Y.~H.; Rastelli,~A.; Schmidt,~O.~G. \emph{Appl Phys Lett}
\textbf{2013},
  \emph{102}, 152105\relax
\mciteBstWouldAddEndPuncttrue
\mciteSetBstMidEndSepPunct{\mcitedefaultmidpunct}
{\mcitedefaultendpunct}{\mcitedefaultseppunct}\relax \EndOfBibitem
\bibitem[Gao \latin{et~al.}(2015)Gao, Imamoglu, Bernien, and Hanson]{Gao2015}
Gao,~W.~B.; Imamoglu,~A.; Bernien,~H.; Hanson,~R. \emph{Nat
Photon}
  \textbf{2015}, \emph{9}, 363--373\relax
\mciteBstWouldAddEndPuncttrue
\mciteSetBstMidEndSepPunct{\mcitedefaultmidpunct}
{\mcitedefaultendpunct}{\mcitedefaultseppunct}\relax \EndOfBibitem
\bibitem[Akopian \latin{et~al.}(2011)Akopian, Wang, Rastelli, Schmidt, and
  Zwiller]{Akopian2011}
Akopian,~N.; Wang,~L.; Rastelli,~A.; Schmidt,~O.~G.; Zwiller,~V.
\emph{Nat
  Photon} \textbf{2011}, \emph{5}, 230 -- 233\relax
\mciteBstWouldAddEndPuncttrue
\mciteSetBstMidEndSepPunct{\mcitedefaultmidpunct}
{\mcitedefaultendpunct}{\mcitedefaultseppunct}\relax \EndOfBibitem
\bibitem[Chekhovich \latin{et~al.}(2012)Chekhovich, Kavokin, Puebla, Krysa,
  Hopkinson, Andreev, Sanchez, Beanland, Skolnick, and
  Tartakovskii]{ChekhovichNatNano2012}
Chekhovich,~E.; Kavokin,~K.; Puebla,~J.; Krysa,~A.; Hopkinson,~M.;
Andreev,~A.;
  Sanchez,~A.; Beanland,~R.; Skolnick,~M.; Tartakovskii,~A. \emph{Nat Nano}
  \textbf{2012}, \emph{7}, 646--650\relax
\mciteBstWouldAddEndPuncttrue
\mciteSetBstMidEndSepPunct{\mcitedefaultmidpunct}
{\mcitedefaultendpunct}{\mcitedefaultseppunct}\relax \EndOfBibitem
\bibitem[Bayer \latin{et~al.}(2002)Bayer, Ortner, Stern, Kuther, Gorbunov,
  Forchel, Hawrylak, Fafard, Hinzer, Reinecke, Walck, Reithmaier, Klopf, and
  Sch\"afer]{Bayer2002}
Bayer,~M.; Ortner,~G.; Stern,~O.; Kuther,~A.; Gorbunov,~A.~A.;
Forchel,~A.;
  Hawrylak,~P.; Fafard,~S.; Hinzer,~K.; Reinecke,~T.~L.; Walck,~S.~N.;
  Reithmaier,~J.~P.; Klopf,~F.; Sch\"afer,~F. \emph{Phys. Rev. B}
  \textbf{2002}, \emph{65}, 195315\relax
\mciteBstWouldAddEndPuncttrue
\mciteSetBstMidEndSepPunct{\mcitedefaultmidpunct}
{\mcitedefaultendpunct}{\mcitedefaultseppunct}\relax \EndOfBibitem
\bibitem[Poem \latin{et~al.}(2010)Poem, Kodriano, Tradonsky, Lindner, Gerardot,
  Petroff, and Gershoni]{Poem2010}
Poem,~E.; Kodriano,~Y.; Tradonsky,~C.; Lindner,~N.~H.;
Gerardot,~B.~D.;
  Petroff,~P.~M.; Gershoni,~D. \emph{Nat Phys} \textbf{2010}, \emph{6}, 993 --
  997\relax
\mciteBstWouldAddEndPuncttrue
\mciteSetBstMidEndSepPunct{\mcitedefaultmidpunct}
{\mcitedefaultendpunct}{\mcitedefaultseppunct}\relax \EndOfBibitem
\bibitem[Puebla \latin{et~al.}(2013)Puebla, Chekhovich, Hopkinson, Senellart,
  Lema\'itre, Skolnick, and Tartakovskii]{Puebla2013}
Puebla,~J.; Chekhovich,~E.~A.; Hopkinson,~M.; Senellart,~P.;
Lema\'itre,~A.;
  Skolnick,~M.~S.; Tartakovskii,~A.~I. \emph{Phys. Rev. B} \textbf{2013},
  \emph{88}, 045306\relax
\mciteBstWouldAddEndPuncttrue
\mciteSetBstMidEndSepPunct{\mcitedefaultmidpunct}
{\mcitedefaultendpunct}{\mcitedefaultseppunct}\relax \EndOfBibitem
\bibitem[Gammon \latin{et~al.}(2001)Gammon, Efros, Kennedy, Rosen, Katzer,
  Park, Brown, Korenev, and Merkulov]{GammonPRL2001}
Gammon,~D.; Efros,~A.~L.; Kennedy,~T.~A.; Rosen,~M.;
Katzer,~D.~S.; Park,~D.;
  Brown,~S.~W.; Korenev,~V.~L.; Merkulov,~I.~A. \emph{Phys. Rev. Lett.}
  \textbf{2001}, \emph{86}, 5176--5179\relax
\mciteBstWouldAddEndPuncttrue
\mciteSetBstMidEndSepPunct{\mcitedefaultmidpunct}
{\mcitedefaultendpunct}{\mcitedefaultseppunct}\relax \EndOfBibitem
\bibitem[Janssens \latin{et~al.}(2001)Janssens, Peeters, and
  Schweigert]{Janssens2001}
Janssens,~K.~L.; Peeters,~F.~M.; Schweigert,~V.~A. \emph{Phys.
Rev. B}
  \textbf{2001}, \emph{63}, 205311\relax
\mciteBstWouldAddEndPuncttrue
\mciteSetBstMidEndSepPunct{\mcitedefaultmidpunct}
{\mcitedefaultendpunct}{\mcitedefaultseppunct}\relax \EndOfBibitem
\bibitem[Mano \latin{et~al.}(2010)Mano, Abbarchi, Kuroda, McSkimming, Ohtake,
  Mitsuishi, and Sakoda]{Mano2010}
Mano,~T.; Abbarchi,~M.; Kuroda,~T.; McSkimming,~B.; Ohtake,~A.;
Mitsuishi,~K.;
  Sakoda,~K. \emph{Applied Physics Express} \textbf{2010}, \emph{3},
  065203\relax
\mciteBstWouldAddEndPuncttrue
\mciteSetBstMidEndSepPunct{\mcitedefaultmidpunct}
{\mcitedefaultendpunct}{\mcitedefaultseppunct}\relax \EndOfBibitem
\bibitem[Snelling \latin{et~al.}(1992)Snelling, Blackwood, McDonagh, Harley,
  and Foxon]{Snelling1992}
Snelling,~M.~J.; Blackwood,~E.; McDonagh,~C.~J.; Harley,~R.~T.;
Foxon,~C. T.~B.
  \emph{Phys. Rev. B} \textbf{1992}, \emph{45}, 3922--3925\relax
\mciteBstWouldAddEndPuncttrue
\mciteSetBstMidEndSepPunct{\mcitedefaultmidpunct}
{\mcitedefaultendpunct}{\mcitedefaultseppunct}\relax \EndOfBibitem
\bibitem[Pryor and Flatt\'e(2006)Pryor, and Flatt\'e]{Pryor2006}
Pryor,~C.~E.; Flatt\'e,~M.~E. \emph{Phys. Rev. Lett.}
\textbf{2006}, \emph{96},
  026804\relax
\mciteBstWouldAddEndPuncttrue
\mciteSetBstMidEndSepPunct{\mcitedefaultmidpunct}
{\mcitedefaultendpunct}{\mcitedefaultseppunct}\relax \EndOfBibitem
\bibitem[Hannak \latin{et~al.}(1995)Hannak, Oestreich, Heberle, Ru¨hle, and
  Ko¨hler]{Hannak1995}
Hannak,~R.~M.; Oestreich,~M.; Heberle,~A.~P.; Ru¨hle,~W.~W.;
Ko¨hler,~K.
  \emph{Solid State Communications} \textbf{1995}, \emph{93}, 313--317\relax
\mciteBstWouldAddEndPuncttrue
\mciteSetBstMidEndSepPunct{\mcitedefaultmidpunct}
{\mcitedefaultendpunct}{\mcitedefaultseppunct}\relax \EndOfBibitem
\bibitem[White \latin{et~al.}(1972)White, Hinchliffe, Dean, and
  Greene]{White1972}
White,~A.; Hinchliffe,~I.; Dean,~P.; Greene,~P. \emph{Solid State
  Communications} \textbf{1972}, \emph{10}, 497 -- 500\relax
\mciteBstWouldAddEndPuncttrue
\mciteSetBstMidEndSepPunct{\mcitedefaultmidpunct}
{\mcitedefaultendpunct}{\mcitedefaultseppunct}\relax \EndOfBibitem
\bibitem[Kosaka \latin{et~al.}(2001)Kosaka, Kiselev, Kim, and
  Yablonovitch]{Kosaka2001}
Kosaka,~H.; Kiselev,~F.~A.,~A. A.~Baron; Kim,~K.~W.;
Yablonovitch,~E.
  \emph{Electronics Letters} \textbf{2001}, \emph{37}, 464--465\relax
\mciteBstWouldAddEndPuncttrue
\mciteSetBstMidEndSepPunct{\mcitedefaultmidpunct}
{\mcitedefaultendpunct}{\mcitedefaultseppunct}\relax \EndOfBibitem
\bibitem[Mej\'{i}a-Salazar \latin{et~al.}(2009)Mej\'{i}a-Salazar,
  Porras-Montenegro, and Oliveira]{MejiaSalazar2009}
Mej\'{i}a-Salazar,~J.~R.; Porras-Montenegro,~N.; Oliveira,~L.~E.
\emph{Journal
  of Physics: Condensed Matter} \textbf{2009}, \emph{21}, 455302\relax
\mciteBstWouldAddEndPuncttrue
\mciteSetBstMidEndSepPunct{\mcitedefaultmidpunct}
{\mcitedefaultendpunct}{\mcitedefaultseppunct}\relax \EndOfBibitem
\bibitem[Jovanov \latin{et~al.}(2011)Jovanov, Eissfeller, Kapfinger, Clark,
  Klotz, Bichler, Keizer, Koenraad, Abstreiter, and Finley]{Jovanov2011}
Jovanov,~V.; Eissfeller,~T.; Kapfinger,~S.; Clark,~E.~C.;
Klotz,~F.;
  Bichler,~M.; Keizer,~J.~G.; Koenraad,~P.~M.; Abstreiter,~G.; Finley,~J.~J.
  \emph{Phys. Rev. B} \textbf{2011}, \emph{83}, 161303\relax
\mciteBstWouldAddEndPuncttrue
\mciteSetBstMidEndSepPunct{\mcitedefaultmidpunct}
{\mcitedefaultendpunct}{\mcitedefaultseppunct}\relax \EndOfBibitem
\bibitem[Bennett \latin{et~al.}(2013)Bennett, Pooley, Cao, Skoeld, Farrer,
  Ritchie, and Shields]{Bennett2013}
Bennett,~A.~J.; Pooley,~M.~A.; Cao,~Y.; Skoeld,~N.; Farrer,~I.;
Ritchie,~D.~A.;
  Shields,~A.~J. \emph{Nat Commun} \textbf{2013}, \emph{4}, 1522\relax
\mciteBstWouldAddEndPuncttrue
\mciteSetBstMidEndSepPunct{\mcitedefaultmidpunct}
{\mcitedefaultendpunct}{\mcitedefaultseppunct}\relax \EndOfBibitem
\bibitem[Nikolaenko \latin{et~al.}(2009)Nikolaenko, Chekhovich, Makhonin,
  Drouzas, Van'kov, Skiba-Szymanska, Skolnick, Senellart, Martrou, Lema\^itre,
  and Tartakovskii]{Nikolaenko2009}
Nikolaenko,~A.~E.; Chekhovich,~E.~A.; Makhonin,~M.~N.;
Drouzas,~I.~W.;
  Van'kov,~A.~B.; Skiba-Szymanska,~J.; Skolnick,~M.~S.; Senellart,~P.;
  Martrou,~D.; Lema\^itre,~A.; Tartakovskii,~A.~I. \emph{Phys. Rev. B}
  \textbf{2009}, \emph{79}, 081303\relax
\mciteBstWouldAddEndPuncttrue
\mciteSetBstMidEndSepPunct{\mcitedefaultmidpunct}
{\mcitedefaultendpunct}{\mcitedefaultseppunct}\relax \EndOfBibitem
\bibitem[Lai \latin{et~al.}(2006)Lai, Maletinsky, Badolato, and
  Imamoglu]{Lai2006}
Lai,~C.~W.; Maletinsky,~P.; Badolato,~A.; Imamoglu,~A. \emph{Phys.
Rev. Lett.}
  \textbf{2006}, \emph{96}, 167403\relax
\mciteBstWouldAddEndPuncttrue
\mciteSetBstMidEndSepPunct{\mcitedefaultmidpunct}
{\mcitedefaultendpunct}{\mcitedefaultseppunct}\relax \EndOfBibitem
\bibitem[Latta \latin{et~al.}(2009)Latta, H\"{o}gele, Zhao, Vamivakas,
  Maletinsky, Kroner, Dreiser, Carusotto, Badolato, Schuh, Wegscheider,
  Atatture, and Imamoglu]{Latta2009}
Latta,~C.; H\"{o}gele,~A.; Zhao,~Y.; Vamivakas,~A.~N.;
Maletinsky,~M.;
  Kroner,~M.; Dreiser,~J.; Carusotto,~I.; Badolato,~A.; Schuh,~D.;
  Wegscheider,~W.; Atatture,~M.; Imamoglu,~A. \emph{Nat Phys} \textbf{2009},
  \emph{5}, 758 -- 763\relax
\mciteBstWouldAddEndPuncttrue
\mciteSetBstMidEndSepPunct{\mcitedefaultmidpunct}
{\mcitedefaultendpunct}{\mcitedefaultseppunct}\relax \EndOfBibitem
\bibitem[Kloeffel \latin{et~al.}(2011)Kloeffel, Dalgarno, Urbaszek, Gerardot,
  Brunner, Petroff, Loss, and Warburton]{Kloeffel2011}
Kloeffel,~C.; Dalgarno,~P.~A.; Urbaszek,~B.; Gerardot,~B.~D.;
Brunner,~D.;
  Petroff,~P.~M.; Loss,~D.; Warburton,~R.~J. \emph{Phys. Rev. Lett.}
  \textbf{2011}, \emph{106}, 046802\relax
\mciteBstWouldAddEndPuncttrue
\mciteSetBstMidEndSepPunct{\mcitedefaultmidpunct}
{\mcitedefaultendpunct}{\mcitedefaultseppunct}\relax \EndOfBibitem
\bibitem[H\"ogele \latin{et~al.}(2012)H\"ogele, Kroner, Latta, Claassen,
  Carusotto, Bulutay, and Imamoglu]{Hoegele2012}
H\"ogele,~A.; Kroner,~M.; Latta,~C.; Claassen,~M.; Carusotto,~I.;
Bulutay,~C.;
  Imamoglu,~A. \emph{Phys. Rev. Lett.} \textbf{2012}, \emph{108}, 197403\relax
\mciteBstWouldAddEndPuncttrue
\mciteSetBstMidEndSepPunct{\mcitedefaultmidpunct}
{\mcitedefaultendpunct}{\mcitedefaultseppunct}\relax \EndOfBibitem
\bibitem[Kessler \latin{et~al.}(2012)Kessler, Reischle, Roßbach, Koroknay,
  Jetter, Schweizer, and Michler]{Kessler2012}
Kessler,~C.~A.; Reischle,~M.; Roßbach,~R.; Koroknay,~E.;
Jetter,~M.;
  Schweizer,~H.; Michler,~P. \emph{physica status solidi (b)} \textbf{2012},
  \emph{249}, 747--751\relax
\mciteBstWouldAddEndPuncttrue
\mciteSetBstMidEndSepPunct{\mcitedefaultmidpunct}
{\mcitedefaultendpunct}{\mcitedefaultseppunct}\relax \EndOfBibitem
\bibitem[Urbaszek \latin{et~al.}(2013)Urbaszek, Marie, Amand, Krebs, Voisin,
  Maletinsky, Hoegele, and Imamoglu]{Urbaszek2013}
Urbaszek,~B.; Marie,~X.; Amand,~T.; Krebs,~O.; Voisin,~P.;
Maletinsky,~P.;
  Hoegele,~A.; Imamoglu,~A. \emph{Rev. Mod. Phys.} \textbf{2013}, \emph{85},
  79--133\relax
\mciteBstWouldAddEndPuncttrue
\mciteSetBstMidEndSepPunct{\mcitedefaultmidpunct}
{\mcitedefaultendpunct}{\mcitedefaultseppunct}\relax \EndOfBibitem
\bibitem[Sundfors(1974)]{Sundfors1974}
Sundfors,~R.~K. \emph{Phys. Rev. B} \textbf{1974}, \emph{10},
4244--4252\relax \mciteBstWouldAddEndPuncttrue
\mciteSetBstMidEndSepPunct{\mcitedefaultmidpunct}
{\mcitedefaultendpunct}{\mcitedefaultseppunct}\relax \EndOfBibitem
\bibitem[Pearson and Faux(2000)Pearson, and Faux]{Pearson2000}
Pearson,~G.~S.; Faux,~D.~A. \emph{Journal of Applied Physics}
\textbf{2000},
  \emph{88}\relax
\mciteBstWouldAddEndPuncttrue
\mciteSetBstMidEndSepPunct{\mcitedefaultmidpunct}
{\mcitedefaultendpunct}{\mcitedefaultseppunct}\relax \EndOfBibitem
\bibitem[Grundmann \latin{et~al.}(1995)Grundmann, Stier, and
  Bimberg]{Grundmann1995}
Grundmann,~M.; Stier,~O.; Bimberg,~D. \emph{Phys. Rev. B}
\textbf{1995},
  \emph{52}, 11969--11981\relax
\mciteBstWouldAddEndPuncttrue
\mciteSetBstMidEndSepPunct{\mcitedefaultmidpunct}
{\mcitedefaultendpunct}{\mcitedefaultseppunct}\relax \EndOfBibitem
\bibitem[Chekhovich \latin{et~al.}(2010)Chekhovich, Makhonin, Skiba-Szymanska,
  Krysa, Kulakovskii, Skolnick, and Tartakovskii]{InPDyn}
Chekhovich,~E.~A.; Makhonin,~M.~N.; Skiba-Szymanska,~J.;
Krysa,~A.~B.;
  Kulakovskii,~V.~D.; Skolnick,~M.~S.; Tartakovskii,~A.~I. \emph{Phys. Rev. B}
  \textbf{2010}, \emph{81}, 245308\relax
\mciteBstWouldAddEndPuncttrue
\mciteSetBstMidEndSepPunct{\mcitedefaultmidpunct}
{\mcitedefaultendpunct}{\mcitedefaultseppunct}\relax \EndOfBibitem
\bibitem[Maletinsky \latin{et~al.}(2009)Maletinsky, Kroner, and
  Imamoglu]{Maletinsky2009}
Maletinsky,~P.; Kroner,~M.; Imamoglu,~A. \emph{Nat Phys}
\textbf{2009},
  \emph{5}, 407 -- 411\relax
\mciteBstWouldAddEndPuncttrue
\mciteSetBstMidEndSepPunct{\mcitedefaultmidpunct}
{\mcitedefaultendpunct}{\mcitedefaultseppunct}\relax \EndOfBibitem
\bibitem[Gammon \latin{et~al.}(1997)Gammon, Brown, Snow, Kennedy, Katzer, and
  Park]{GammonScience}
Gammon,~D.; Brown,~S.~W.; Snow,~E.~S.; Kennedy,~T.~A.;
Katzer,~D.~S.; Park,~D.
  \emph{Science} \textbf{1997}, \emph{277}, 85--88\relax
\mciteBstWouldAddEndPuncttrue
\mciteSetBstMidEndSepPunct{\mcitedefaultmidpunct}
{\mcitedefaultendpunct}{\mcitedefaultseppunct}\relax \EndOfBibitem
\bibitem[Makhonin \latin{et~al.}(2010)Makhonin, Chekhovich, Senellart,
  Lema\'itre, Skolnick, and Tartakovskii]{MakhoninPRB}
Makhonin,~M.~N.; Chekhovich,~E.~A.; Senellart,~P.; Lema\'itre,~A.;
  Skolnick,~M.~S.; Tartakovskii,~A.~I. \emph{Phys. Rev. B} \textbf{2010},
  \emph{82}, 161309\relax
\mciteBstWouldAddEndPuncttrue
\mciteSetBstMidEndSepPunct{\mcitedefaultmidpunct}
{\mcitedefaultendpunct}{\mcitedefaultseppunct}\relax \EndOfBibitem
\end{mcitethebibliography}

\providecommand{\latin}[1]{#1}
\providecommand*\mcitethebibliography{\thebibliography} \csname
@ifundefined\endcsname{endmcitethebibliography}
  {\let\endmcitethebibliography\endthebibliography}{}

\end{document}